\documentclass[apj,numberedappendix]{emulateapj}
\usepackage{graphics,epsf}
\usepackage{amsmath}                
\usepackage{amsfonts}               
\usepackage{amssymb}                
\usepackage{epsfig}
 \usepackage{epstopdf}
\usepackage{multirow}
\usepackage[para,online,flushleft]{threeparttable}

\def \kev{\rm{keV}}

\def \s{~\rm{s}}
\def \km{~\rm{km}}

\def \erg{~\rm{erg}}

\def \yr{~\rm{yr}}

\def \pc{~\rm{pc}}
\def \kpc{~\rm{kpc}}

\def \kyr{~\rm{kyr}}

\def \keV{~\rm{keV}}

\def \GHz{~\rm{GHz}}
\def \MHz{~\rm{MHz}}
\newcommand{\nar}{{~\rm New Astronomy Reviews}}
\newcommand{\na}{{~\rm New Astronomy}}
\newcommand{\pasa}{{~\rm Publications of the Astronomical Society of Australia}}
\newcommand{\jcap}{{~\rm Cosmology Astropart. Phys.}}

\begin{document}

\title{Neutron star natal kick and jets in core collapse supernovae}

\author{Ealeal Bear\altaffilmark{1} and Noam Soker\altaffilmark{1}}

\altaffiltext{1}{Department of Physics, Technion -- Israel Institute of Technology, Haifa
32000, Israel; ealealbh@gmail.com; soker@physics.technion.ac.il}

\begin{abstract}
We measure the angle between the neutron star (NS) natal kick direction and the inferred direction of jets according to the morphology of 12 core collapse supernova remnants (SNR), and find that the distribution is almost random, but missing small angles.
 The 12 SNRs are those for which we could both identify morphological features that we can attribute to jets and for which the direction of the NS natal kick is given in the literature. Unlike some claims for spin-kick alignment, here we rule out jet-kick alignment. We discuss the cumulative distribution function of the jet-kick angles under the assumption that dense clumps that are ejected by the explosion accelerate the NS by the gravitational attraction, and suggest that the jet feedback explosion mechanism might in principle account for the distribution of jet-kick angles.
\end{abstract}


\section{INTROCUTION}
\label{sec:intro}

Many core collapse supernovae (CCSNe) leave behind a neutron star (NS) remnant that is born with a significant non-zero velocity, called natal kick velocity, with typical values of $200 - 500\km \s^{-1}$ and up to about $1000\km \s^{-1}$
(e.g., \citealt{Cordesetal1993,LyneLorimer1994, Chatterjeeetal2005}).
These values are larger than what can be accounted for by the disruption of a close binary system. Therefore, it is likely that an asymmetrical explosion mechanism is the cause of the natal kick velocity (e.g., \citealt{Laietal2006, Wongwathanaratetal2013}; for recent summary of many studies on the natal kick see \citealt{Janka2017}).
In a recent through study, \cite{Katsudaetal2018} find from X-ray measurements of six
supernova remnants (SNRs) that elements between silicon and calcium are generally ejected opposite to the direction of NS motion. This, they argue, supports the connection of NS natal kick to asymmetrical explosion.

Other mechanisms that have been proposed in the past, cannot work.
Asymmetric neutrino emission by itself
cannot account for the observed kick velocities  (e.g., \citealt{Lai2003, Wongwathanaratetal2010, Nordhausetal2010, Nordhausetal2012, Katsudaetal2018} and references therein).
Scenarios that are based on momentum imparted by asymmetrical two opposite jets cannot explain the high natal kick velocities as they require massive jets, and hence, as argued by, e.g., \cite{Nordhausetal2012}, they require rapid pre-collapse core rotation, and therefore, this scenario might be at best viable for a small portion of natal kick cases. Possible combinations of these scenarios have also been raised. For example, the combination of magnetic fields and rapid rotation which can cause jets that might induce a kick (e.g., see discussion by \citealt{Wangetal2006}).
As the source of the momentum of the NS is the two asymmetrical opposite jets, according to this mechanism the jets' axis (defined as the line along the directions of the two opposite jets) and kick direction tend to be aligned. This is in contradiction with the results we present in the present study. In what follow we will not consider these and other mechanisms (e.g., \citealt{CharbonneauZhitnitsky2010}), and we will refer only to asymmetrical explosion mechanisms that impart momentum to the newly born NS.

Many observational and theoretical papers study and discuss the relation between the spin and kick directions (e.g., \citealt{SpruitPhinney1998,FryerKusenko2006, NgRomani2007, Wangetal2007}).
In the Crab nebula \citep{Kaplanetal2008} and the Vela nebula \citep{Laietal2001} observations imply an almost alignment between the NS kick direction and the spin direction. While some papers find a strong correlation between the kick and the spin directions (e.g., \citealt{Dodsonetal2003, Johnstonetal2005, Johnstonetal2006}), other papers, such as \cite{BrayEldridge2016}, find no statistical preference for the kick orientation. \cite{NgRomani2006} find that spin-kick angle in the pulsar of the Crab nebula is $26^\circ$ rather than the previously determined angle of $8^\circ$ (also \citealt{Wangetal2007}).

In a recent study \cite{HollandAshfordetal2017} compare both the directions and magnitudes of the NS kick velocities with the asymmetrical geometry of SNRs. They look at the dipole, quadrupole, and octupole power-ratios of the SNR morphologies, and find no correlation of SNR asymmetry with the magnitude of the kick velocity. They do find that the NS kick directions are preferentially opposite to the bulk of the X-ray emission.

In the present study we compare kick directions with another geometrical property of the SNRs.
We examine the relation between the kick direction and the line connecting the two opposite ears of SNR, or other morphological features that hint at jets. We follow \cite{GrichenerSoker2017} and generally define ears as two opposite protrusions from the main SNR shell. We further take the view that the ears were shaped by jets launched from the newly born NS during the explosion of the SN \citep{GrichenerSoker2017, Bearetal2017}. The ears' axis is defined as the line connecting the tips of the two ears. Hence, from here on we will refer to ears' axis and jets' axis as meaning the same, but keep in mind that what we observe in the SNRs are the ears.

 It is not clear if the jets we study here can leave a mark during the SN phase itself. \cite{Piranetal2018} attribute the excess of high velocity material in hydrogen-stripped CCSNe to relativistic choked jets that accelerated material to high velocities. The jets we discuss here might be weaker and the SN main shell more massive. From the typical ears we observe during the SNR phase, we can estimate that the velocity of the ears is only $\approx 10-20 \%$ higher than the main shell. We also do not expect the ears to change the luminosity as they cover a small area and the emissivity of the ears will not differ much from that of the main shell. However, a more detailed study should be conducted to answer this question.  

 The motivation to our study that focuses on the relation between the direction of ears' (or jets') axis and the direction of the natal kick is our view that in all cases with ears the explosion was driven by jets.
 The ears are shaped by the last jet-launching episode because these jets are launched just after the previous jets have expelled the inner core. Therefore, the last jets might flow freely to the edge of the expanding envelope, and gently breakout, leaving the imprint of two opposite ears (similar to the modeling of Cassiopeia~A by \citealt{Orlandoetal2016}). The energy of the jets that inflated the ears is only a fraction of the explosion energy because the explosion was driven by several earlier jet-launching episodes \citep{Bearetal2017}.

There is no need for the pre-collapse core of the exploding star to have fast rotation, or even a mild rotation as in the model of \cite{Wheeleretal2002} for jets, as convective regions in the pre-collapse core \citep{GilkisSoker2014, GilkisSoker2016} and/or instabilities in the shocked zones around the newly born NS \citep{Papishetal2015} can supply stochastic angular momentum to the gas that is accreted on to the NS.
Most pronounced of these instabilities that might supply stochastic angular momentum are the spiral modes of the standing accretion shock instability (SASI; on the spiral SASI modes see, e.g., \citealt{BlondinMezzacappa2007, Rantsiouetal2011, Fernandez2015, Kazeronietal2017}).
If the accreted gas launches jets, then because of its stochastic angular momentum the jets' axis will change its direction over time. This is termed the jittering jets explosion mechanism
\citep{PapishSoker2011, PapishSoker2014}.
If the pre-collapse core is rapidly rotating, then the jets will maintain a more or less constant axis.
In both cases, the jets operate in a negative feedback mechanism (see review by \citealt{Soker2016Rev}).
We adopt here the view that the jet feedback explosion mechanism can account for all CCSNe, from typical energies of about $10^{51} \erg$ \citep{PapishSoker2011} and up to super-energetic (or superluminous) CCSNe, even when a magnetar is formed (e.g., \citealt{Soker2016Mag1, Chenetal2017, SokerGilkis2017}).

We construct our paper as follows. In section \ref{sec:angle} we discuss each of the 12 SNRs for which we could both identify ears (or another morphological feature that hint at jets) in available images and find the kick direction in the literature. The important new result in that section is the collection of the 12 projected angles between the kick direction and the direction of the ears' (jets') axis in the 12 SNRs. These angles are summarized in section \ref{subsec:sample}. We discuss each SNR in more detail in section \ref{subsec:detail}. Readers who are interested only in the results and their analysis can skip section \ref{subsec:detail}.  In section \ref{sec:analysis} we analyze the distribution of these angles and compare it with two distributions, a random distribution and a distribution which assumes that the kick and jets' axis are perpendicular to each other. We discuss the results in the frame of the jets feedback explosion mechanism.
We present our short summary in section \ref{sec:summary}.

\section{THE ANGLES BETWEEN KICK DIRECTION AND JETS AXIS}
\label{sec:angle}
\subsection{Sample and measured angles}
\label{subsec:sample}

In this section we review 12 SNRs for which we found in the literature both morphological features that we can identify with jets and the direction of the motion of their central NS.
We list the SNRs and the name of their NSs in the first and second columns of Table \ref{Table1}, respectively.
We measured the angle $\alpha$ between the direction of the NS natal kick and the line along the directions of the two opposite jets, which we term the jets' axis. We list the values of $\alpha$ in the third column, and the source for the assumed jets' axis in the fourth column of Table \ref{Table1}.
Because in some cases the two ears are not exactly on opposite sides of the center and/or in some cases one or two of the ears do not possess exact symmetry around an axis, we cannot always determine accurate jets' axis direction. We estimate that these departures from pure axi-symmetry lead to general uncertainties in the values of $\alpha$ for the different SNRs that are about several degrees, e.g., about $\pm 5^\circ$.
 When available, we also list the angle $\phi$ between the NS spin and the kick direction (fifth column), and the references for that value (sixth column).
\begin{table}
  \begin{threeparttable}
\centering
\caption{Angles between the jets' axis and the NS kick direction}
\label{Table1}
\begin{tabular}{llllll}
\hline
SNR         & PSR            & $\alpha$ & Jets & $\phi$ & Spin \\
\hline
Cassiopeia A &               & 88       & G    &   &         \\
Puppis A    & PSR J0821−4300 & 40       & G    &   &          \\
RCW 103     & 1E 161348−5055 & 80       & B    &   &          \\
PKS 1209-51 & 1E 1207.4-5209 & 54       & Here &   &           \\
CTB 109     & 1E 2259+586    & 42       & Here &   &      \\
S147        & PSR J0538+2817 & 40       & G    & 12 & NR  \\
G292.0+1.8  & PSR J1124–5916 & 70       & B    &22, 70& W,P \\
Vela        & B0833-45       & 30       & G    & 10 &NR \\
G327.1-1.1  &                & 45       & Here &      &        \\
3C58        & PSR J0205+6449 & 60       & G   & 21 & NR \\
Crab        & PSR B0531+21   & 18       & G   & 26 & NR \\
W44         & PSR B1853+01   & 15       & G   &   &  \\   \hline
\end{tabular}
\begin{tablenotes}
      \small
      \item The first and second columns list the name of the SNR and the NS, respectively. The angle $\alpha$ (in degrees) is our measured angle between what we take as the jets' axis and the NS kick direction. The fourth column lists the source for the jets' axis: G: \cite{GrichenerSoker2017}; B: \cite{Bearetal2017}; Here: this study. The angle $\phi$ (in degrees) is the angle between the NS kick direction and the NS spin for which the references are given in the last column: NR: \cite{NgRomani2007}; W: \cite{Wangetal2006}; P: \cite{Parketal2007}.
    \end{tablenotes}
  \end{threeparttable}
\end{table}

Morphological features that we identify with jets are mainly two opposite ears (defined in section \ref{sec:intro}) and two opposite bright arcs. The identification of jets with ears follows our earlier papers, and it is based on the morphologies of planetary nebulae with ears and similar structures that are attributed to jets (\citealt{Bearetal2017, BearSoker2017, GrichenerSoker2017}). As well, \cite{TsebrenkoSoker2013} demonstrated that jets can form ears in SNRs of Type Ia SNe. The flow that leads to ears in remnants of CCSNe is somewhat different than that in Type Ia SNe. The last jets to be launched by the exploding massive star carry a small, but non-negligible energy of the main supernova shell. Each jet pushes its way from inside and leaves a mark on the outskirts of the SNR \citep{TsebrenkoSoker2013}. If the jets are stronger, they can penetrate throughout the shell and form a morphology like in RCW 103 \citep{Bearetal2017}.  The jets' axis is taken to be along the line connecting the two opposite ears or along the arcs. For 9 SNRs we take the direction of the jets from previous papers, as listed in the fourth column of Table \ref{Table1}.
For 3 other SNRs we assume here the axis of the two opposite jets.

In the present study we are concerned only with the morphologies of the ears and other features that indicate jets. The relative brightness of the ears and the main SNR shell might depend on local conditions that include the intensity and morphology of the magnetic field lines, the population of high energy electrons, and clumps that result from the CSM or ISM. The magnetic fields and high energy electrons determine the X-ray and radio synchrotron emission. Thermal X-ray emission and the population of high energy electrons depend on shocks, that in turn depend also on dense clumps. But neither of these factors that determine the emission will change in any significant manner the morphology of the ears. Only a massive CSM or ISM medium can do that. 

  In section \ref{subsec:detail} we describe each SNR in more detail. Readers who are interested only in the results and their analysis can skip section \ref{subsec:detail} and go directly to the analysis in section \ref{sec:analysis}.

\subsection{Detailed description of SNRs}
\label{subsec:detail}

  In the figures to follow we draw both the jets' axis and the NS natal kick direction in the upper panel for each of the 12 SNRs. From there we calculated the angle between the kick direction and jets' axis, as listed in the third column of Table \ref{Table1}. Other panels in the figures to follow are intended to show the NS natal kick direction and the jets' direction as taken from the literature. We turn to describe in short each SNR and its basic properties that might be relevant to the analysis.

\begin{figure}[ht!]
\centering
\includegraphics[trim= 0.0cm 0.0cm 0.0cm 0.0cm,clip=true,width=0.8\textwidth]{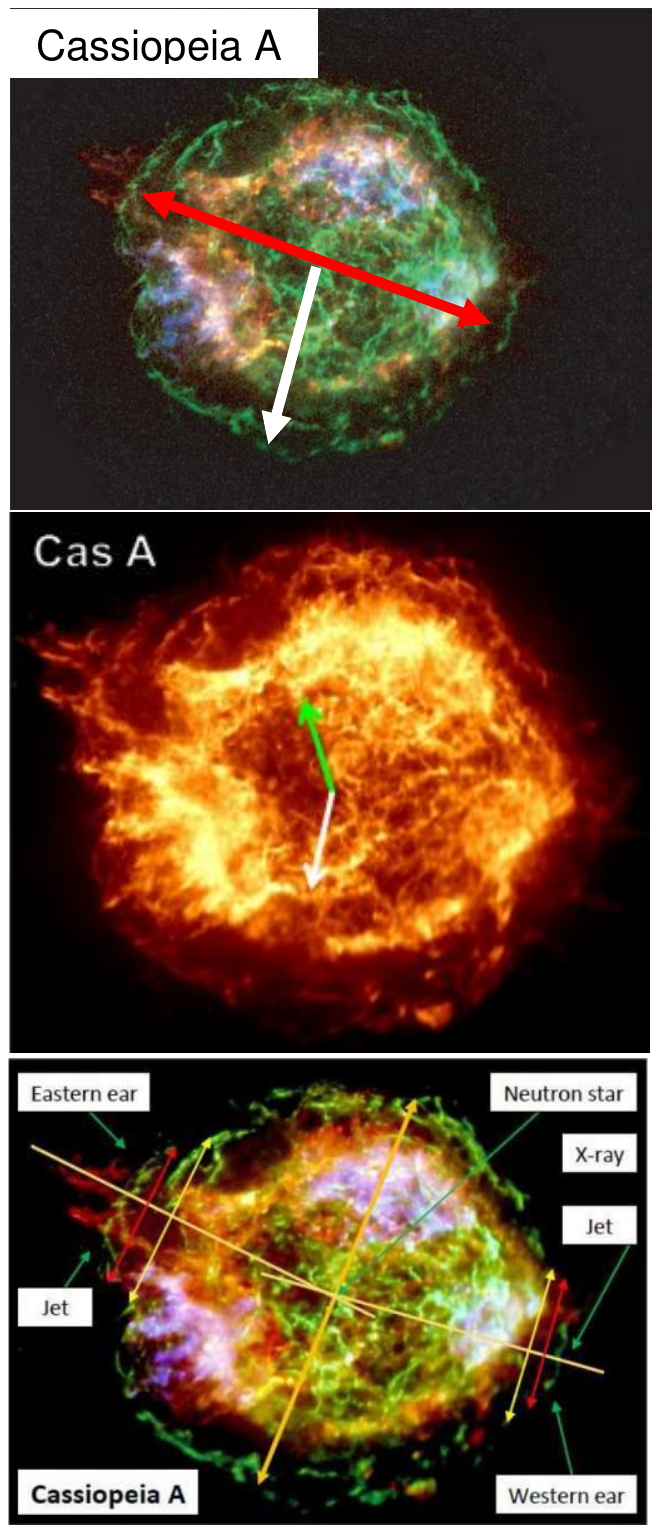}
\vskip -6.0 cm
\caption{The upper panel is an X-ray image of SNR Cassiopeia A \citep{Hwangetal2004}. The image shown is a three-color image of Cas A with red = Si He$\alpha$ (1.78–2.0 \kev), blue = Fe K (6.52–6.95 \kev), and green$= 4.2–6.4 \kev$ continuum. Si-rich ejecta in red is in the northeast direction and Fe-rich faint ejecta in blue is in the southeast direction (see also \citealt{Hughesetal2000, Hwangetal2000}). The white arrow points in the direction of the NS kick, as we take from the middle panel. The middle panel is a $0.5-2.1 \kev$ Chandra and ROSAT image, where the green arrow points from the explosion site to the direction of the dipole moment and the white arrow points in the direction of NS motion (taken from \citealt{HollandAshfordetal2017}). We mark the jets' axis by the red double-headed arrow. It is taken to be the line connecting the two ears as marked by \cite{GrichenerSoker2017} in the lower panel. }
  \label{fig:CassA}
\end{figure}

\textit{Cassiopeia A (Cas~A, 3C 461, G111.7-2.1).} Cas~A is at a distance of $3.4 \kpc$ (e.g., \citealt{Reedetal1995}). The mass of the progenitor prior to the explosion could have reached $20M_\odot$ (e.g., \citealt{Willingaleetal2003}), and its age is assumed to be $330\yr$ (e.g., \citealt{Yakovlevetal2011}).
It resulted from an asymmetric type IIb explosion (e.g., \citealt{Krauseetal2008}). Jets have previously been modeled for Cas~A (e.g., \citealt{Schureetal2008}). One of the outcomes from their model is that jets can accompany the explosion even if the SNR appears spherically symmetric.
 \cite{DeLaneySatterfield2013} estimate the proper motion of the NS star as $V_{\rm NS} =390 \pm 400 \km\s^{-1}$.
The upper panel in Fig. \ref{fig:CassA} is an X-ray image taken from \cite{Hwangetal2004}, where the white arrow points in the direction of NS motion taken from \cite{HollandAshfordetal2017} as presented in the middle panel. The red double-headed arrow in the upper panel is along the direction of the two opposite jets taken from \cite{GrichenerSoker2017} as presented in the lower panel.

\begin{figure}[ht!]
\centering
\includegraphics[trim= 0.0cm 0.0cm 0.0cm 0.0cm,clip=true,width=0.7\textwidth]{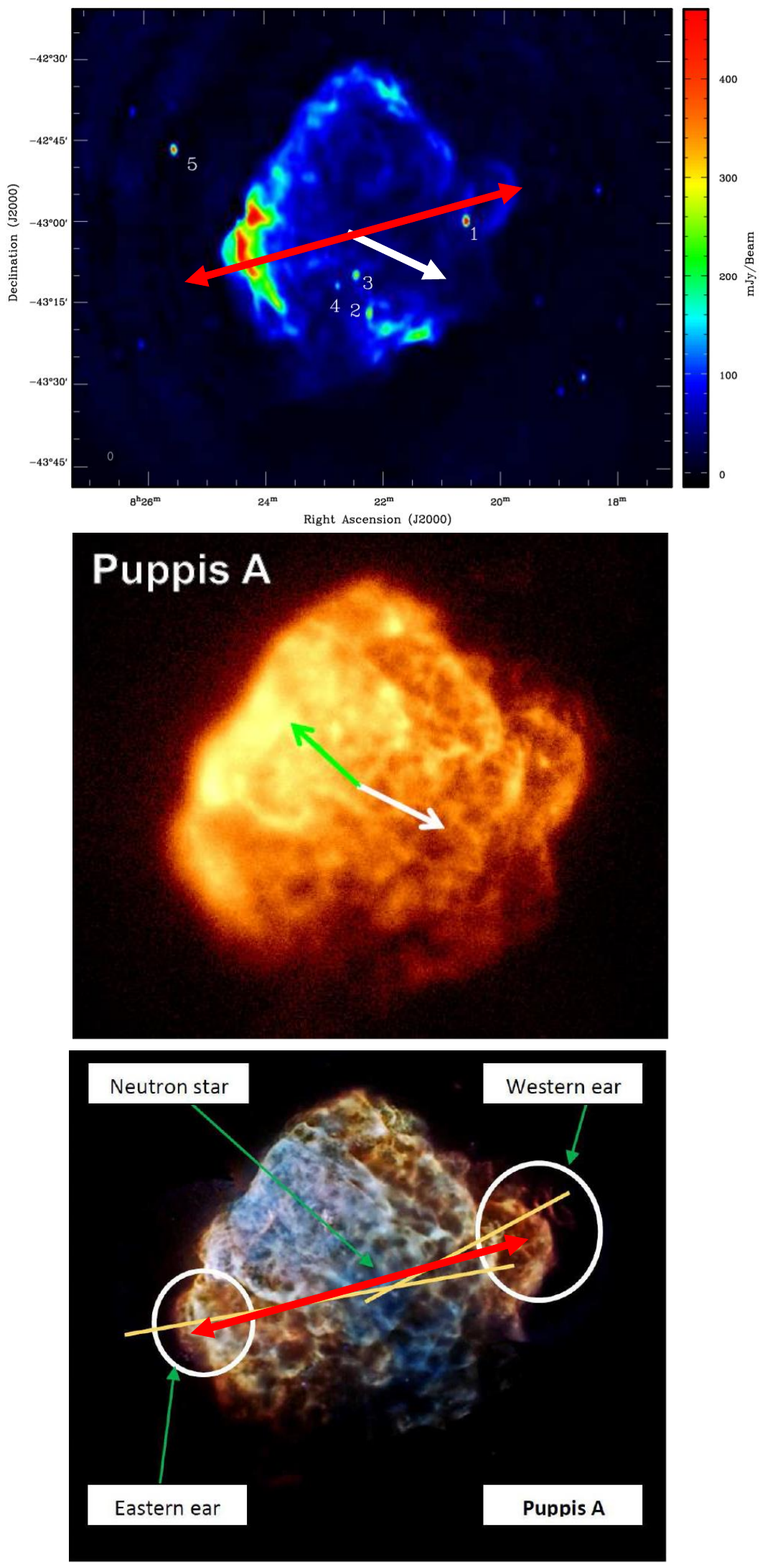}
\vskip -2.0 cm
\caption{The upper panel is a 1.4~$\GHz$ radio image of SNR Puppis A taken from \cite{ReynosoWalsh2015}, to which we added the jets' axis and kick direction. The flux density scale is shown at the right.
The NS kick direction is marked in a white arrow (upper panel), based on the middle panel (\citealt{HollandAshfordetal2017}. Details for this panel are the same as in Fig. \ref{fig:CassA}.)
The lower panel is taken from \cite{GrichenerSoker2017} and gives a full view of Puppis A in X-ray, where red, green, and blue correspond to the 0.3-0.7, 0.7-1.0, and $1.0-8.0 \kev$ bands, respectively (from \citealt{Dubneretal2013}).
}
  \label{fig:PuppisA}
\end{figure}

\textit{Puppis A (G260.4-03.4).} Its age is estimated as ranging from $3700 - 4450 \yr$  (e.g., \citealt{Beckeretal2012}).
Jets have already been proposed to be the shaping mechanism of this SNR (e.g., \citealt{Castellettietal2006}). Furthermore, \cite{Reynosoetal2003} claim that the morphological features of this SNR (e.g., the alignment between optical expansion center and the lobes) are caused by jets.
The NS (called RX J0822−4300) transverse motion is measured at  $1570 \pm  240 \km \s^{-1}$ towards the west-southwest, assuming a distance of $2 \kpc$ \citep{WinklerPetre2007}. We draw the jets' axis and the kick direction in the upper panel of Fig. \ref{fig:PuppisA}. The NS motion is taken from \cite{HollandAshfordetal2017} as shown in the middle panel, and the jets' axis is taken from \cite{GrichenerSoker2017} as shown in the lower panel.

\begin{figure}[ht!]
\centering
\includegraphics[trim= 0.0cm 0.0cm 0.0cm 0.0cm,clip=true,width=0.60\textwidth]{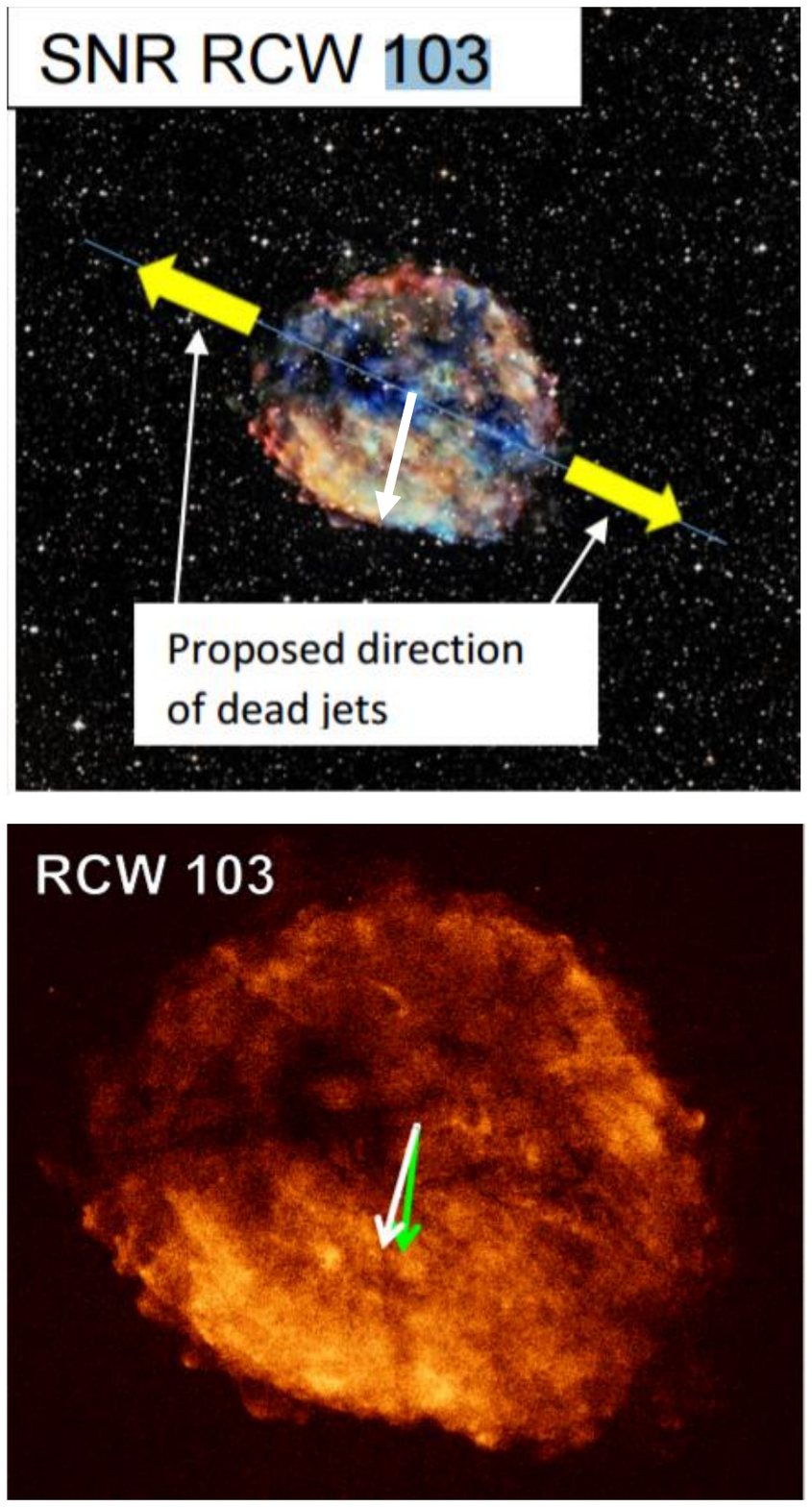}
\vskip -3.0 cm
\caption{The upper panel is an X-ray image of RCW~103 in three energy bands (low=red, medium=green, highest=blue) combined with an optical image from the Digitized Sky Survey. The original image is from the Chandra website and it is based on \cite{Reaetal2016}, while the yellow arrows that depict the direction of the jets were added by \cite{Bearetal2017}.
A white arrow is the NS kick direction, copied from the lower panel that is taken from \cite{HollandAshfordetal2017}. Arrows in the lower panel are as in Fig. \ref{fig:CassA}.}
  \label{fig:RCW103}
\end{figure}

\textit{RCW 103 (G332.4-00.4).} On the upper panel in Fig. \ref {fig:RCW103} (taken from  \citealt{Bearetal2017} and based upon \citealt{Reaetal2016}), we mark the proposed jets' axis in yellow arrows. The NS motion is marked in a white arrow taken from \citep{HollandAshfordetal2017} as noted in the lower panel. Although there are no ears in this SNR, in a previous paper (\citealt{Bearetal2017}; see figure 3 there) we have compared the morphology of this SNR to several planetary nebulae and from that deduced the direction of the jets that have shaped this SNR. RCW~103 estimated age is $\approx 2000\yr$ (e.g., \citealt{Carteretal1997}) and its estimated distance is $\approx 3.3 \kpc$ (e.g., \citealt{Reynosoetal2004, Xingetal2014} and references therein). The kick velocity of the NS (1E~16134825055) is estimated to be $\approx 810 - 1300 \km \s^{-1}$ (for more details see \citealt{Toriietal1998}).

\begin{figure}[ht!]
\centering
\includegraphics[trim= 0.0cm 0.0cm 0.0cm 0.0cm,clip=true,width=0.70\textwidth]{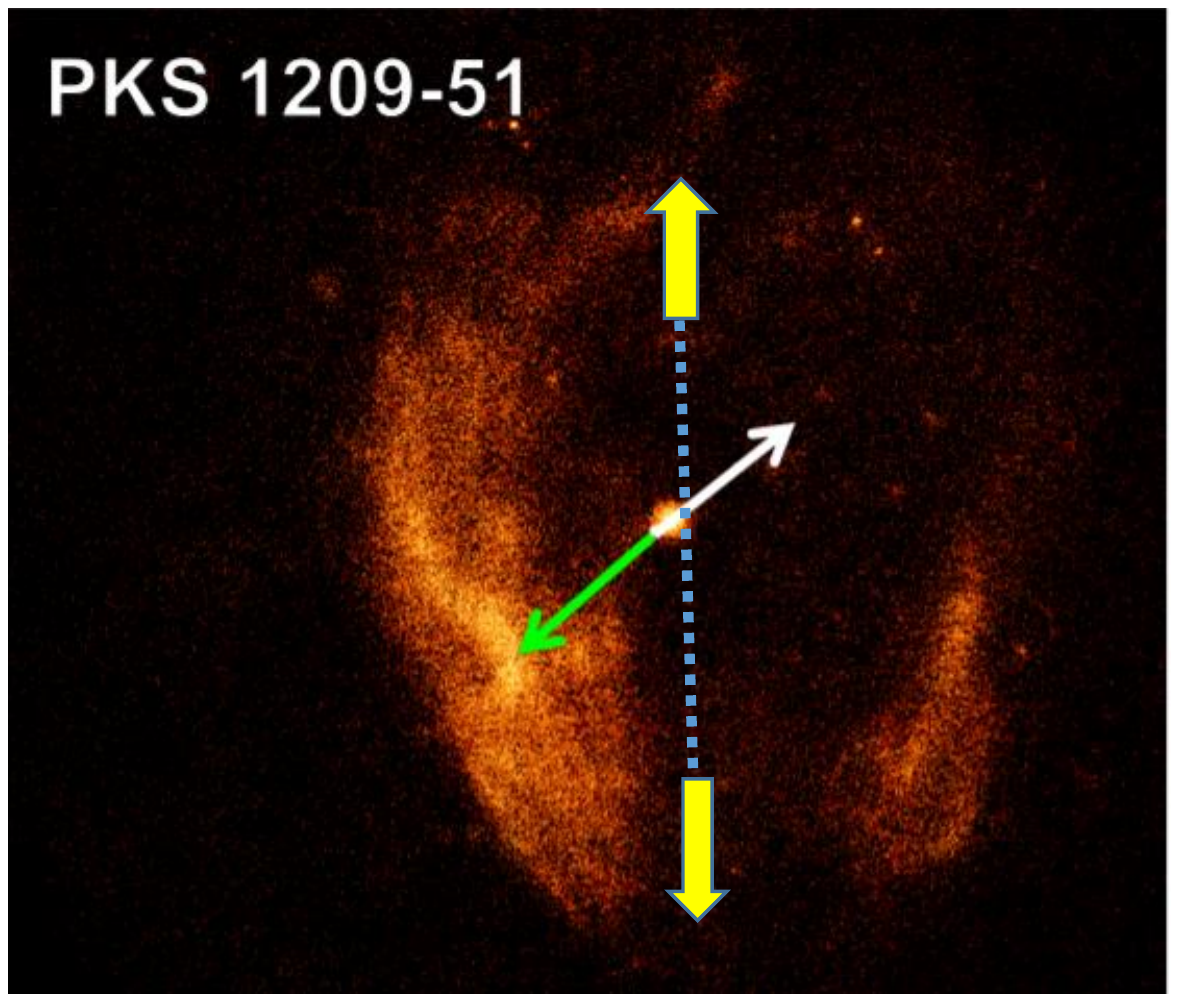}
\vskip -9.0 cm
\caption{An image of SNR PKS1209-51 with the NS direction of motion marked by a white arrow (taken from \citealt{HollandAshfordetal2017}). Green arrow is as in Fig. \ref{fig:CassA}.
The yellow arrows present our proposed direction of the two jets that shaped this SNR during the explosion.}
  \label{fig:PKS1209}
\end{figure}

\textit{PKS 1209-51/52 (G296.510.0).} This SNR is at a distance of $\approx 2.1 \kpc$ (e.g., \citealt{Giacanietal2000}) and its age is estimated to be $\approx 7000\yr$ (e.g., \citealt{Pavlovetal2002}). We take the NS kick direction from \cite{HollandAshfordetal2017} and mark it with a white arrow on Fig. \ref{fig:PKS1209}.
Similar to RCW103, this SNR has no clear ears, and we propose that the jets that shaped this SNR where launched along a direction between the two bright arcs, as we mark by yellow arrows connected by a cyan-dotted line in Fig. \ref{fig:PKS1209}.

\begin{figure}[!t]
\centering
\includegraphics[trim= 0.0cm 0.0cm 0.0cm 0.0cm,clip=true,width=0.60\textwidth]{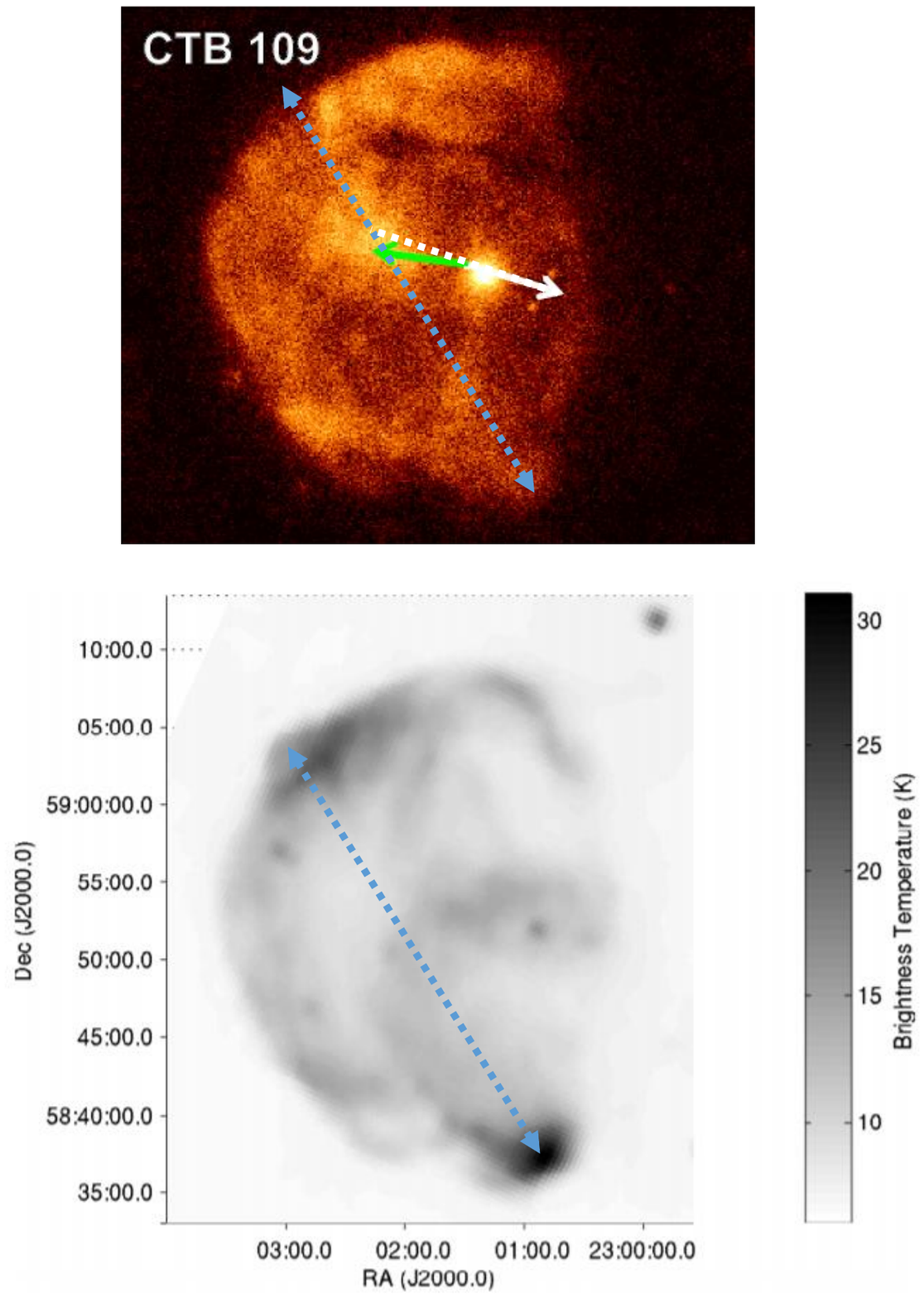}
\vskip -3.0 cm
\caption{Images of the SNR CTB~109. The upper panel is a $0.5 - 2.1 \keV$ Chandra and ROSAT image of CTB 109 with a white arrow that marks the NS motion (taken from \citealt{HollandAshfordetal2017}; the green arrow points from the explosion site to the direction of the dipole moment).
The lower panel is a $1420 \MHz$ radio continuum image, taken from \cite{Bolteetal2015}
and based on \cite{Kothesetal2002} from the Canadian Galactic Plane Survey (CGPS, \citealt{Tayloretal2003}).
We added a double-dotted cyan arrow on the two panels to mark our proposed jets' axis. }
  \label{fig:CTB109}
\end{figure}

\textit{CTB 109 (G109.1-01.0).} CTB~109 is a radio and X-ray bright shell-type SNR at a distance of $\approx 3.2 \kpc$ (e.g., \citealt{KothesFoster2012, SanchezCrucesetal2018}). We take its image together with a white arrow that marks the direction of motion of the NS from \citep{HollandAshfordetal2017} and present it as the upper panel of Fig. \ref{fig:CTB109}. The morphology of these ears is not exactly as observed in some other SNRs. They are very bright in the radio, as presented in the lower panel of Fig. \ref{fig:CTB109}. We take the line connecting the two bright ears to be the jets' axis, and mark our proposed jet direction with a dotted cyan double-headed arrow on the two panels of Fig. \ref{fig:CTB109}.

\begin{figure}[ht!]
\centering
\includegraphics[trim= 0.0cm 0.0cm 0.0cm 0.0cm,clip=true,width=0.50\textwidth]{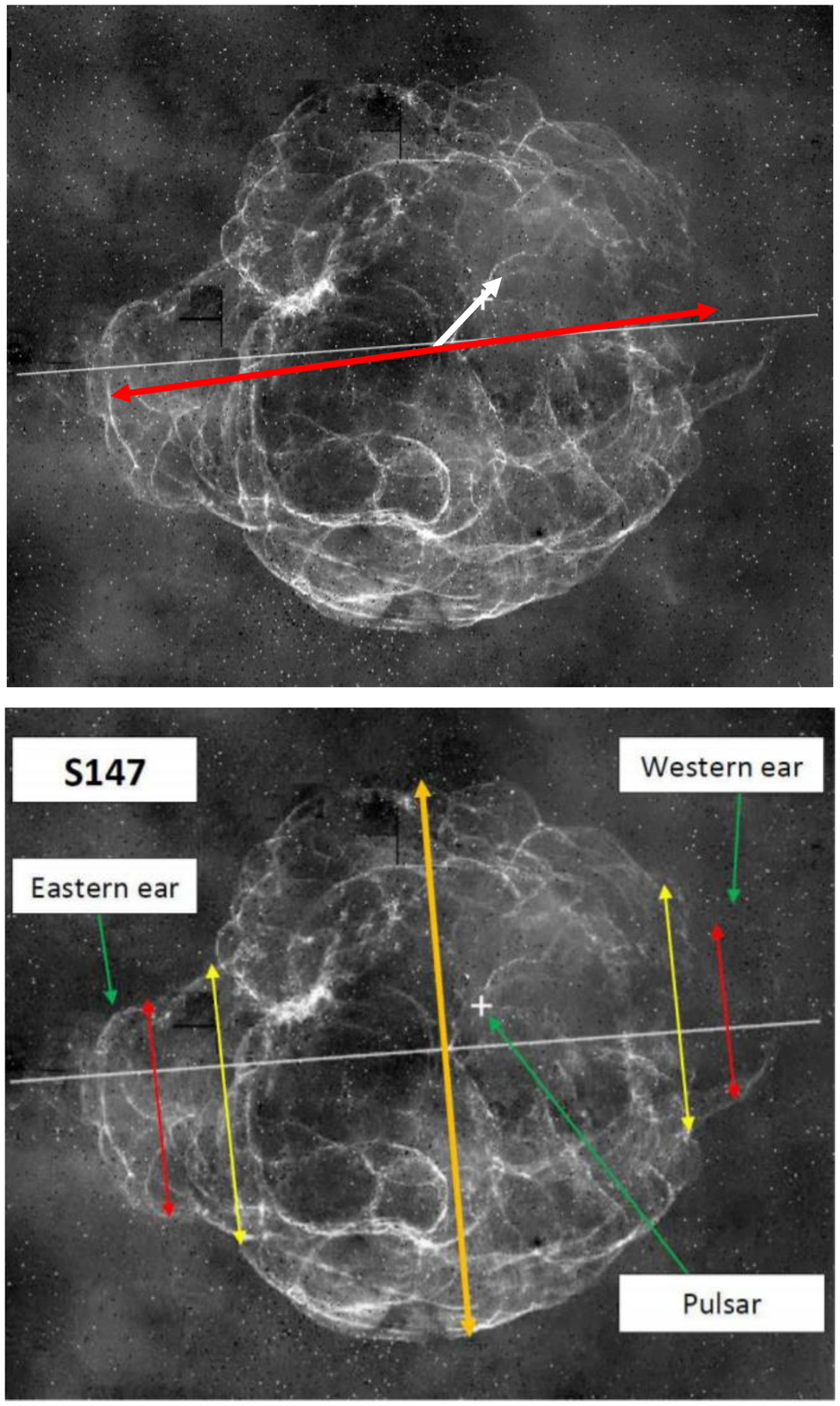}
\vskip -1.0 cm
\caption{The upper panel is an H$\alpha$ image of the supernova remnant S~147 taken from \cite{Gvaramadze2006} based on \cite{Drewetal2005}. We added a white arrow to indicate the NS (pulsar PSR~J0538+2817) motion according to \citep{Gvaramadze2006}, from the center of the SNR towards the NS (white plus sign). The white line drawn in the east-west direction shows the bilateral symmetry axis of the SNR (for more details see \citealt{Gvaramadze2006}).
The jets' axis that we mark by the double-headed red arrow on the upper panel is based on the lower panel taken from \cite{GrichenerSoker2017}. }
  \label{fig:S147}
\end{figure}

\textit{S~147 (G180.0−1.7).} Its distance is estimated as $1.47 \kpc$, its age is taken to be $20 -100 \kyr$, and the spin-kick angle is $12 ^\circ$ (e.g., \citealt{Ngetal2007, Romani2005}).
The kick velocity of the NS (PSR J0538+2817) of S147 is estimated as $\approx 800 \km \s^{-1}$ (e.g., \citealt{RomaniNg2003}).
The upper panel in Fig. \ref{fig:S147} is taken from \cite{Gvaramadze2006} based on \cite{Drewetal2005}.
We added the white arrow to mark the NS motion as reported by \cite{Gvaramadze2006}. It is consistent with the direction from the geometric center of S 147 to the present position of the pulsar as marked by a white plus sign. We mark the jets' direction according to the lower panel taken from \cite{GrichenerSoker2017}.

\begin{figure}[ht!]
\centering
\includegraphics[trim= 0.0cm 0.0cm 0.0cm 0.0cm,clip=true,width=0.80\textwidth]{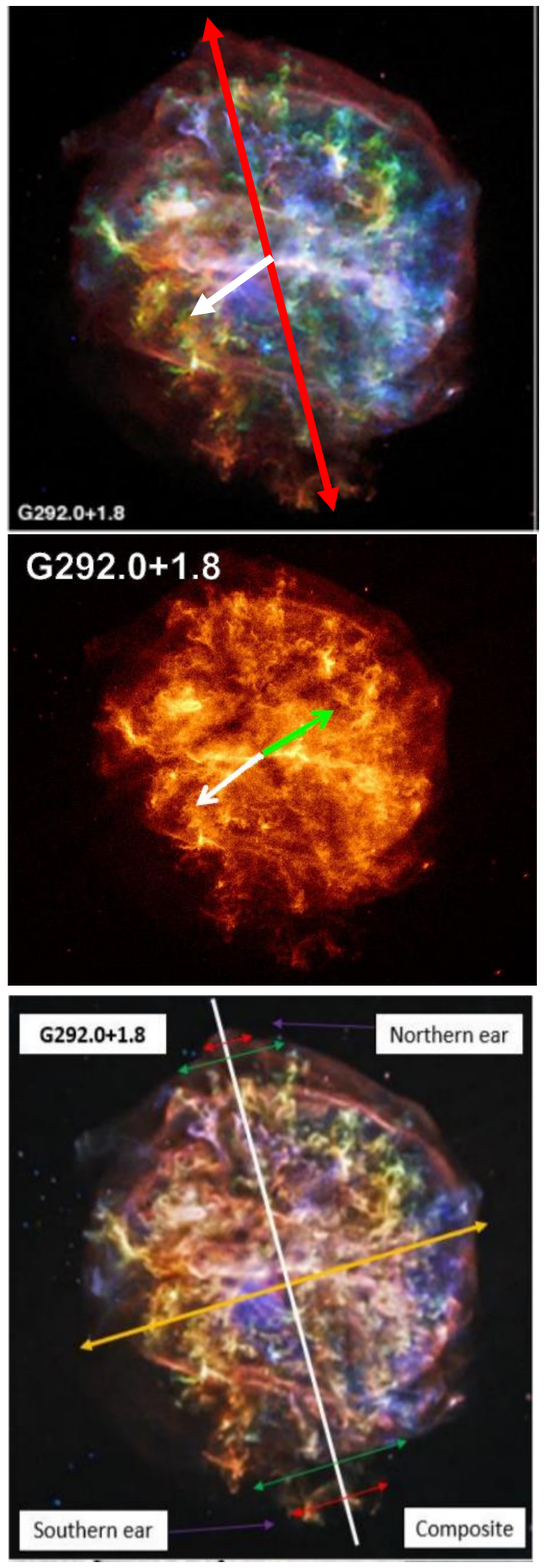}
\vskip -2.1 cm
\caption{The upper panel is a composite image of G292.0+1.8 taken from the Chandra gallery and based on \cite{Parketal2007}. Red, orange, green and blue colors represent different X-ray lines, while white represents the optical band.
The middle panel is taken from \cite{HollandAshfordetal2017}. We copied the white arrow that represents the NS motion to the upper panel. Green arrow is as in Fig. \ref{fig:CassA}. The lower panel is taken form \textbf{{{{\cite{Bearetal2017}}}}} to indicate the jets' axis between the protrusions. We copied the jets' axis to the upper panel (red double-headed arrow).}
  \label{fig:G292_0}
\end{figure}

\textit{G292.0+1.8.} G292.0+1.8 is a Galactic oxygen-rich CCSNR (e.g., \citealt{Bhaleraoetal2015}). Its pulsar J1124−5916 is apparently off the geometric center of the SNR and with an estimated velocity of $770\km\s^{-1}$, a distance of $4.8 \kpc$, and an age of $1660 \yr$ (e.g., \citealt{Hughesetal2001} and references therein). \cite{Parketal2007} suggest that the angle between the spin and the kick direction can be $70^\circ$ or less. Others also point to a misalignment but derive much smaller angles, e.g., $22^\circ$ \citep{Wangetal2006}. On the upper panel of fig. \ref{fig:G292_0} we mark the NS motion (white arrow) copied from the middle panel taken from \cite{HollandAshfordetal2017}, and the jets' axis (double-headed red arrow) based on the lower panel taken from \cite{GrichenerSoker2017}.

\begin{figure}[ht!]
\centering
\includegraphics[trim= 0.0cm 0.0cm 0.0cm 0.0cm,clip=true,width=0.90\textwidth]{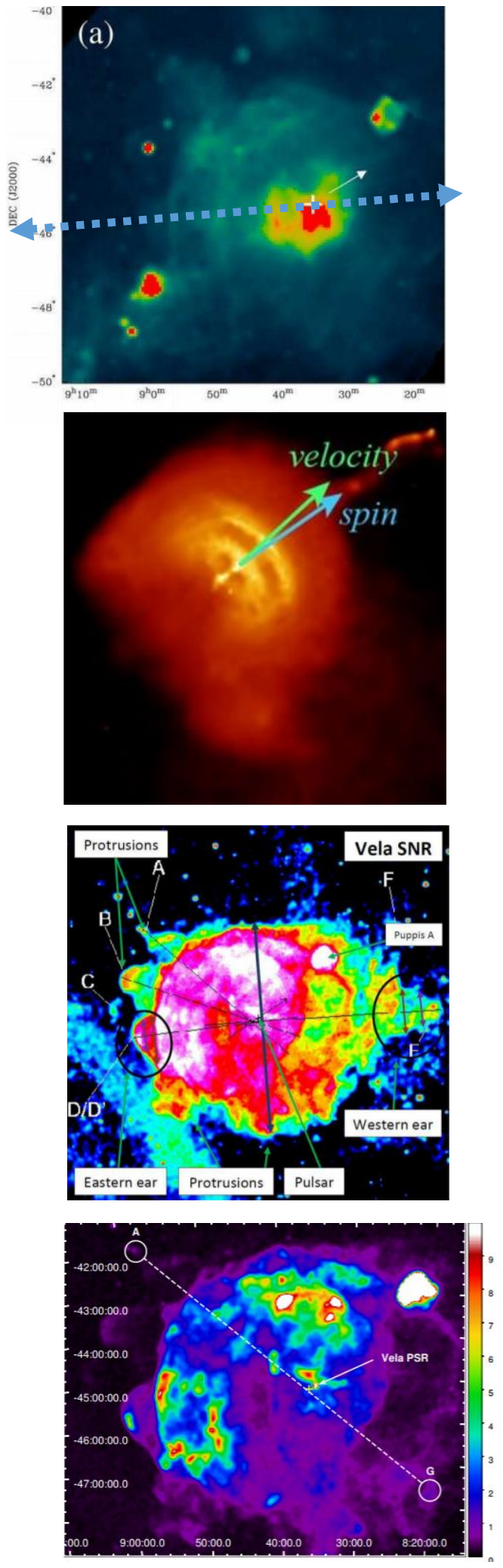}
\vskip -2.5 cm
\caption{The upper panel is a $2.4 \GHz$ radio image of the Vela SNR taken from \cite{GaenslerSlane2006} and based on \cite{Duncanetal1996}. The cross indicates the location of the associated pulsar B0833–45, while the white arrow indicates its direction of motion.
The second panel is taken from the Max Planck Institute for Radio astronomy newsletter \citep{Pavlovetal2008,Noutsosetal2012}.
The third panel is the Vela SNR taken from \cite{GrichenerSoker2017} where the proposed jets' axis is marked as a line connecting the ears. It is a ROSAT all-sky survey image (0.1 - 2.4 \kev) taken from \cite{Aschenbachetal1995}.
We mark the jet direction on the upper panel according to the third panel \citep{GrichenerSoker2017}, so the angle between the jets' axis and the NS motion will be clearer. The fourth panel is a recent observation of Vela which suggests a different jets' axis \citep{Garciaetal2017}.}
  \label{fig:Vela}
\end{figure}

\textit{Vela (G263.9-03.0).} Vela is at a distance of $\approx 350-500\pc$ (e.g., \citealt{Micelietal2008, Aschenbachetal1995} respectively) and at an age of $\approx 10^4\yr$ (e.g., \citealt{Micelietal2008}).  The progenitor mass is estimated as $\approx 15 M_\odot$ (e.g., \citealt{ChenGehrels1999}).
The angle between the NS spin and kick direction is considered to be aligned at $10^\circ$ (e.g. \citealt{Pavlovetal2001, NgRomani2007}).
\cite{Garciaetal2017} analyze two opposite Si-rich knots in Vela, and argue that they were ejected by jets. The direction of the axis of their suggested two opposite jets is almost perpendicular to the NS kick velocity, and is different than what we take here to be the jets' axis. Such a case might be the outcome of the jittering jets explosion mechanism (see section \ref{sec:intro}). The two double-jets were launched at two different times out of several jets'-launching episodes \citep{PapishSoker2011}.
The two upper panels in Fig. \ref{fig:Vela} focus on the NS (pulsar B0833–45) and its direction of motion. The two lower panels indicate possible jet directions, taken from \cite{GrichenerSoker2017} and \cite{Garciaetal2017}, respectively.
We assume that the jets' axis is as in the third panel \citep{GrichenerSoker2017}. Taken the jets' axis from the fourth panel as suggested by \cite{Garciaetal2017} would give a larger value of $\alpha$.

\begin{figure}[ht!]
\centering
\includegraphics[trim= 0.0cm 0.0cm 0.0cm 0.0cm,clip=true,width=0.60\textwidth]{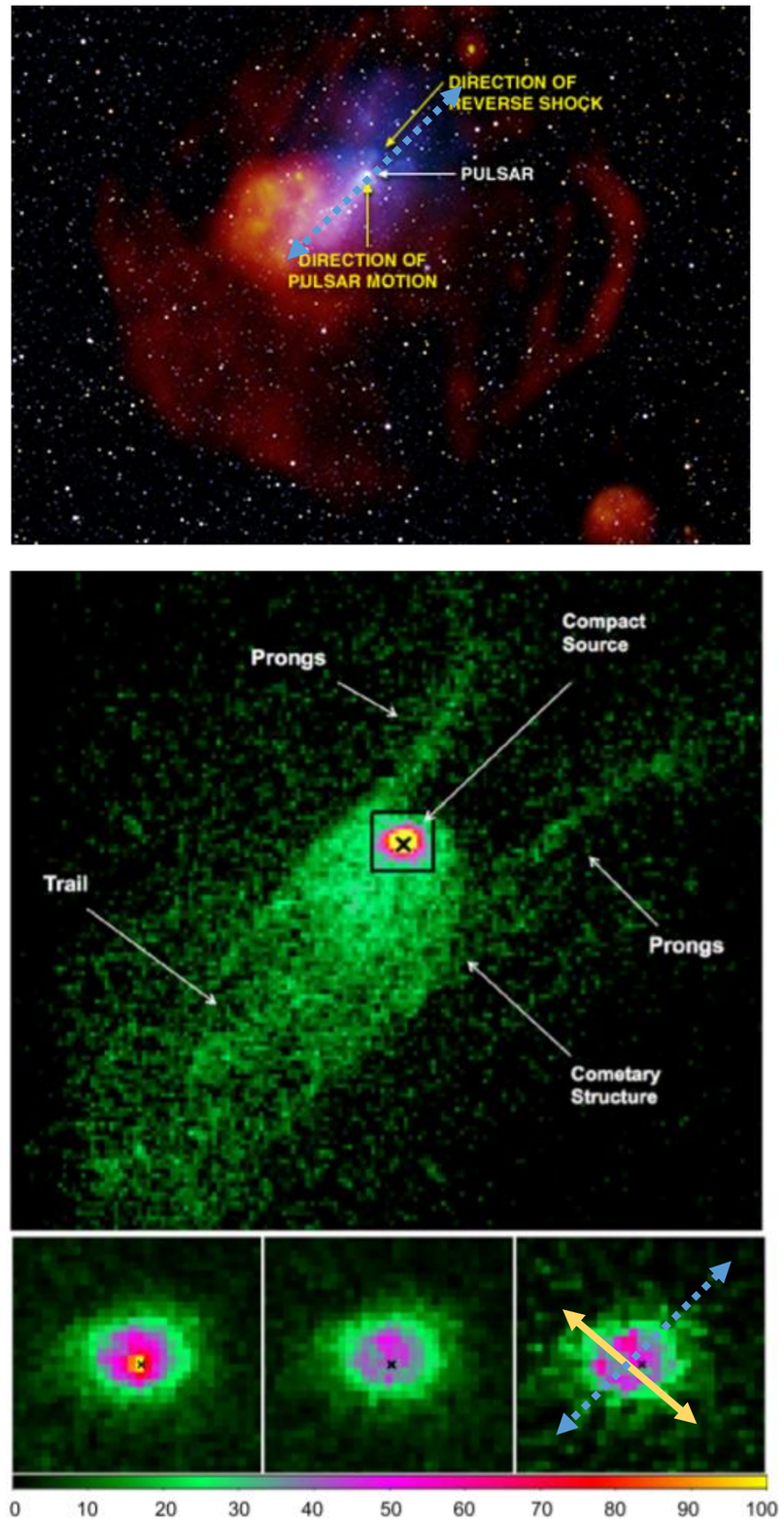}
\vskip -1.0 cm
\caption{The upper panel is a composite image of G327.1-1.1 (from Chandra website based on \citealt{Temimetal2009}; blue: X-ray; red: radio-MOST; yellow: radio-ATCA; RGB: infrared).
The yellow arrow (upper panel) represents the NS kick direction (from the Chandra website).
The lower panels are from \cite{Temimetal2015}. We mark with a double-headed yellow arrow what we identify as the plane of the torus, and in the upper panel we mark with a cyan dotted double-headed arrow our assumed jets' axis. }
  \label{fig:G327_1}
\end{figure}

\textit{G327.1-1.1.} Its estimated age is $\approx 11000- 29000\yr$ depending on the model that is used (e.g., \citealt{Temimetal2009} and references therein). The NS direction of motion is marked (in the original figure) in the upper panel of Fig. \ref{fig:G327_1} by a yellow arrow (taken from the Chandra Gallery, based on \citealt{Temimetal2009}).
We identify no ears in this SNR. However, \cite{Temimetal2015} identify a torus that is seen in the small lower-right panel of Fig. \ref{fig:G327_1}. Based on its similarity to bright tori in other pulsar wind nebulae (e.g., \citealt{KargaltsevPavlov2008}), we draw by a double-headed yellow arrow the plane of the torus on that image.
We take the jets' axis to be perpendicular to the torus, as drawn on the upper panel of Fig. \ref{fig:G327_1} with a cyan dashed double-headed arrow.

\begin{figure}[ht!]
\centering
\includegraphics[trim= 0.0cm 0.0cm 0.0cm 0.0cm,clip=true,width=0.55\textwidth]{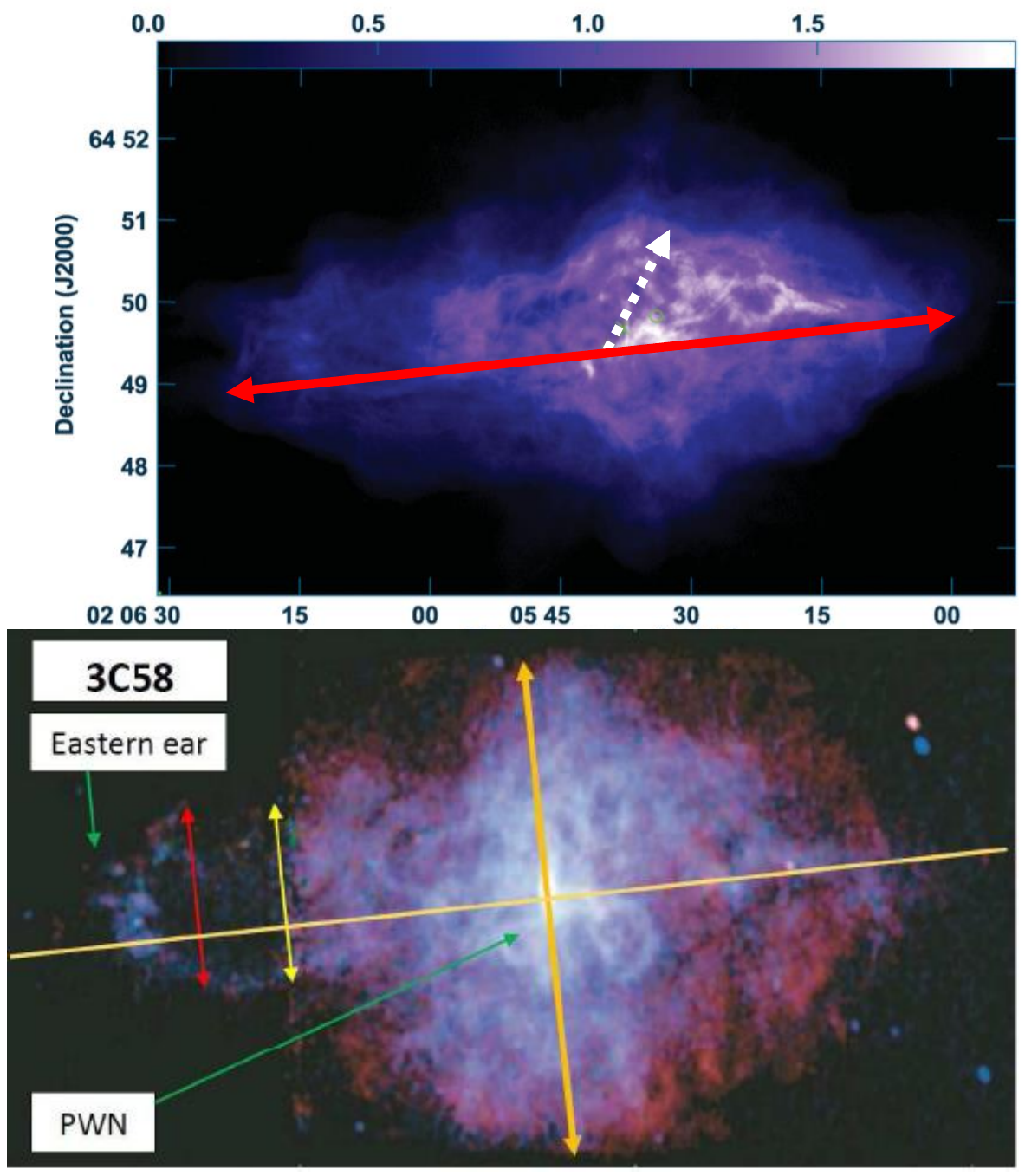}
\vskip -5.0 cm
\caption{The upper panel is a $1.4 \GHz$ VLA radio image of 3C58 taken from  \cite{Bietenholzetal2013}.
We mark with a white dotted-arrow the NS motion according to \cite{Bietenholzetal2013}.
We mark the proposed jets' axis with a red double-headed arrow according to the jets' axis in the lower panel taken from \cite{GrichenerSoker2017}, who made the marks on an ACIS/Chandra X-ray images of 3C58 that is based on the work of \cite{Slaneetal2004}.}
  \label{fig:3C58}
\end{figure}

\textit{3C58 (G130.7+03.1).} It is at a distance of $\approx 2 \kpc$ with an estimated age of $\approx 830 \yr$ (e.g., \citealt{Kothes2013}). \cite{Slaneetal2004} discuss the jet morphology of this SNR, but they focus on the curved features of the jet.
\cite{NgRomani2007} measured the angle between the spin and the kick direction of the NS (PSR J0205+6449) to be $21^\circ$. The NS motion as we marked it on the upper panel of Fig. \ref{fig:3C58} is according to \cite{Bietenholzetal2013}. The lower panel shows the jets' axis as was marked by  \cite{GrichenerSoker2017}.

\begin{figure}[ht!]
\centering
\includegraphics[trim= 0.0cm 0.0cm 0.0cm 0.0cm,clip=true,width=0.70\textwidth]{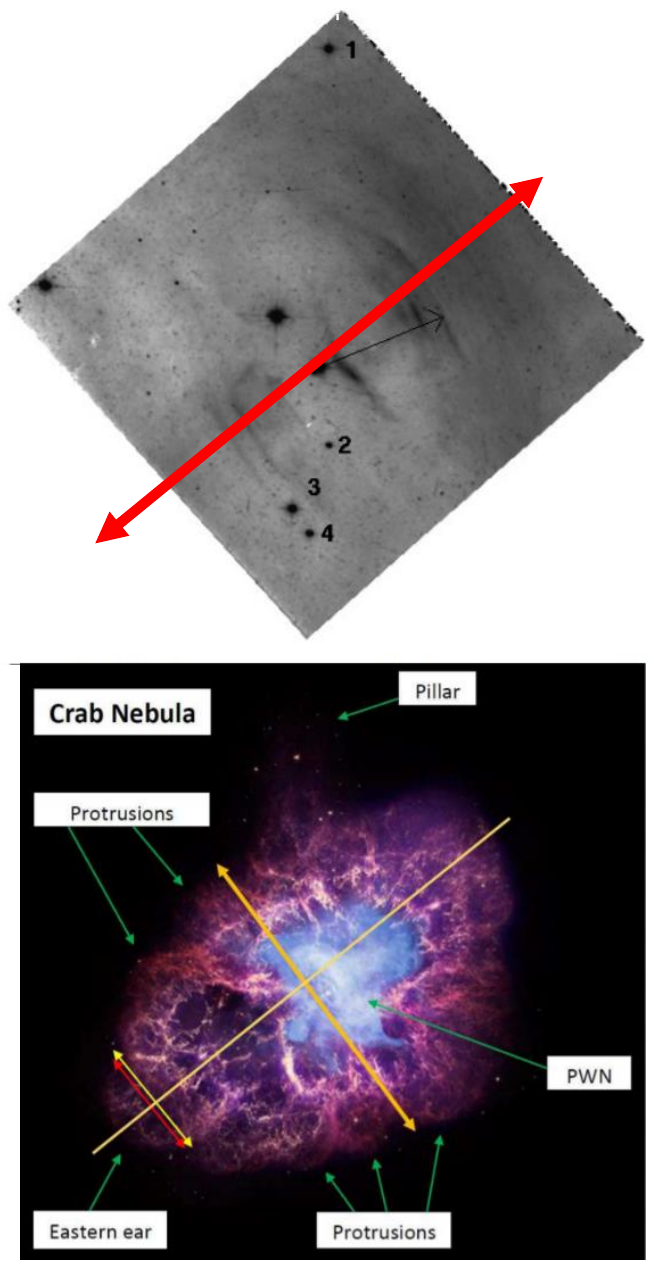}
\vskip -7.0 cm
\caption{The upper panel is an HST image of the inner Crab Nebula together with an arrow in the direction of motion of the Crab pulsar as marked by \cite{CaraveoMignani1999}. The numbers 1 to 4 are common reference stars.
In the upper panel we added the proposed jets' axis with a red double-headed arrow according to the lower panel taken from \cite{GrichenerSoker2017}.
The background image in the lower panel is a composite image from Chandra`s gallery assembled from X-ray (blue; \citealt{Sewardetal2006}), optical (red-yellow; \citealt{Hester2008}) and IR (purple; NASA/JPL-Caltech/Univ). }
  \label{fig:Crab}
\end{figure}

\textit{The Crab (G184.6-05.8).} It was formed by either a Type II or a Type Ib SN (e.g., \citealt{PolcaroMartocchia2006}) that exploded in 1054. The upper panel of Fig. \ref{fig:Crab} is taken from \cite{CaraveoMignani1999} where they marked the direction of the NS (PSR B0531+21) by a black arrow.
The lower panel shows the jets' axis as was marked by \cite{GrichenerSoker2017}, that we copied as a red double-headed arrow to the upper panel.
As discussed by \cite{Wangetal2007}, the spin-kick angle of the crab pulsar B0531+21 has previously been considered to be aligned ($8^\circ$) but now the angle is estimated to be $26^\circ$ (e.g., \citealt{NgRomani2006}).

\begin{figure}[ht!]
\centering
\includegraphics[trim= 0.0cm 0.0cm 0.0cm 0.0cm,clip=true,width=0.50\textwidth]{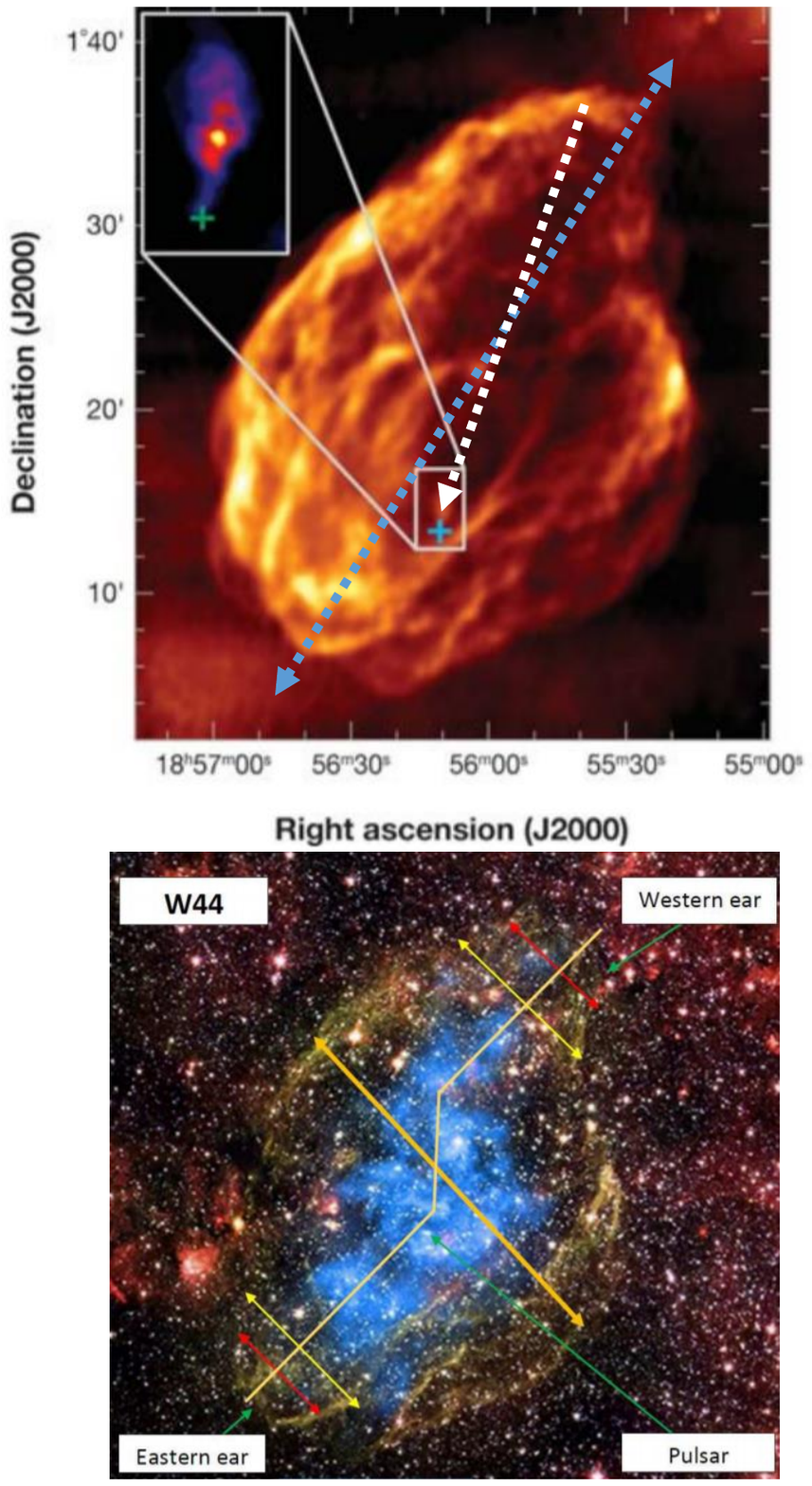}
\vskip -1.0 cm
\caption{The upper panel is a $1.4 \GHz$ VLA image of W44 taken from \cite{GaenslerSlane2006} and based on \cite{Giacanietal1997}. The inset in the upper panel shows an $8.4 \GHz$ VLA data on the region surrounding the associated young pulsar B1853+01 based on \cite{Frailetal1996}.
The position of the pulsar B1853+01 is marked by a cross.
We mark the general direction of the NS kick by a dashed white arrow. We take this direction to be opposite to the synchrotron trail.
We added a double-dotted cyan arrow that connects the two ears according to the lower panel taken from \cite{GrichenerSoker2017}.
The lower panel is a composite image taken from the Chandra gallery and with marks added by \cite{GrichenerSoker2017}.
 The cyan represents X-ray (based on \cite{Sheltoneteal2004}), while the red, blue and green represent infra-red (based on NASA/JPLCaltech).
}
  \label{fig:W44}
\end{figure}

\textit{W44 (G034.6-00.5).} The age and distance of W44 are estimated to be $\approx 20,000\yr$ and $\approx 3.1 \kpc$, respectively (e.g., \citealt{Cardilloetal2014} and references therein). The direction of the NS motion is estimated according to the inset in the upper panel of Fig. \ref{fig:W44} (taken from \citealt{GaenslerSlane2006}). As noted by \cite{Frailetal1996} the synchrotron trail points in a northwest direction which is opposite to the direction of the NS, and this supports their contention that the pulsar originated close to the geometric center of W44. We mark the general direction of the NS motion with a white dotted arrow in the upper panel. \cite{GrichenerSoker2017} marked the two jets to be in opposite directions but not along the same line, as we show in the lower panel of Fig. \ref{fig:W44}. We take the jets' axis to be the line connecting the two ears as we mark by a cyan double-dotted-arrow in the upper panel.

\section{ANALYSIS}
\label{sec:analysis}

\begin{figure}[!t]
\centering
\vskip -2.00cm
\hskip -2.00cm
\includegraphics[trim= 0.0cm 0.0cm 0.0cm 0.0cm,clip=true,width=0.60\textwidth]{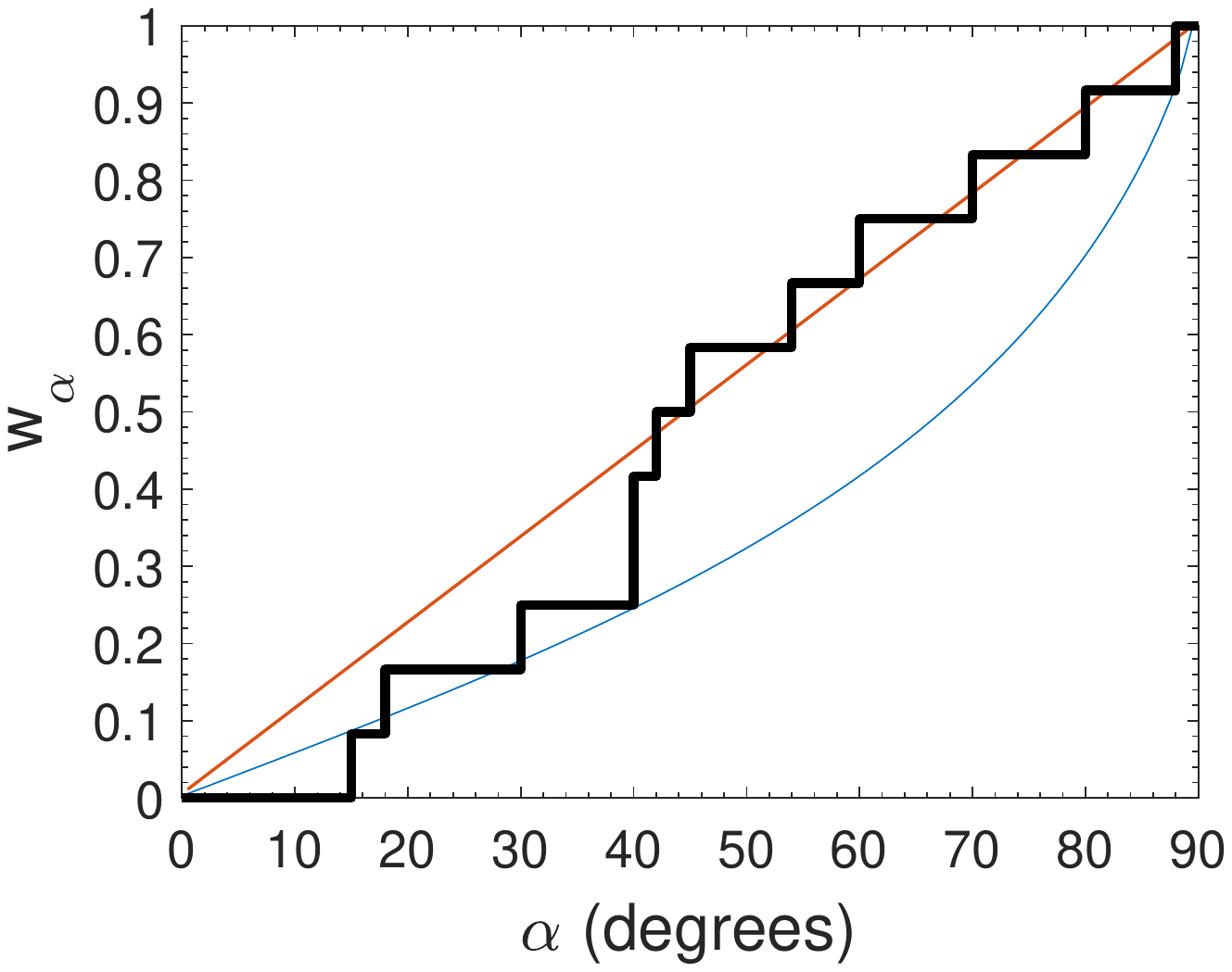}
\vskip -3.9cm
\caption{The cumulative distribution function ${\rm W}_{\alpha}$ of projected jets-kick angles for the 12 observed objects (black line). The straight orange line is the expected random cumulative distribution function, while the convex blue line is the expected cumulative distribution function when in all cases the NS kick velocity is perpendicular to the jets' axis.
}
  \label{fig:distribution}
\end{figure}

In Fig. \ref{fig:distribution} we present the cumulative distribution function of the projected angle $\alpha$ between the NS kick direction and the jets' axis. We recall that we assume that the ears are formed by jets, and take the direction of each ear as the direction of a jet that inflated the ear (see section \ref{sec:intro}). In some SNRs that have no ears we take the jets' axis to be along the two opposite bright arcs. The straight orange line on Fig. \ref{fig:distribution} depicts the expected distribution for a random angle (no correlation) between the SN kick and jets directions, while the convex blue line represents the expected distribution when for all objects the NS kick is perpendicular to the jets' symmetry axis.

The equation for the convex blue line is derived by projecting the two perpendicular lines (those of the jets' direction and of the kick direction) onto the plain of the sky, giving each possible orientation in space the appropriate weight. Let $0 \le \theta \le \pi$ be the angle between the kick direction and the line of sight. The direction of the jets' axis is in the plane perpendicular to the kick direction. Let $0 \le \phi < \pi$ be the angle of the jets' axis in that plane, where $\beta=0$ corresponds to the case when the jets' direction is just behind the kick direction. The relative weight of this position is $2 \sin \theta ~ d \theta ~d \phi$. The projected angle on the sky between the kick and jets' axis is given by $\tan \alpha = \tan \phi/ \cos \theta$. Numerically integrating over all possible values of $\theta$ and $\phi$ with the appropriate weight, gives the distribution for the perpendicular case.

We performed a Kolmogorov-Smirnov test for the compatibility of the sample of 12 objects with the two distributions.
We find the maximum distance on the graph between the observed and expected random distributions to be $D=0.2$. From this we calculate $P=0.67$,
namely, there is a chance of $67 \%$ that the 12 objects are compatible with the random distribution (straight line).
For the compatibility with the perpendicular distribution (lower blue line) we find $D=0.33$ from which we calculate $P=0.12$. 
Namely, we can reject the perpendicular distribution with $88 \%$ confident.
We raise below a third possibility.

  Before we raise this third possibility, we must emphasize in the strongest possible way that we obtain this distribution from only 12 objects. Therefore, there are very large uncertainties in how the real distribution should look like. With many more objects it might turn out to be a random distribution, or else, less likely, it might turn out to be more like the perpendicular distribution. Below, we simply assume, with all the caution we can apply, that the cumulative distribution function we find here is close to the real one. The basic feature in the cumulative distribution function is that relative to a random distribution systems are missing for angles of $\alpha \la 15^\circ$.

 This is the place to reemphasize that while most previous studies of the kick direction in CCSNe have assumed that the explosion is driven by neutrinos, basically the delayed neutrino mechanism (e.g. \citealt{Mulller2016}, for a recent review), we adopt the jet feedback explosion mechanism (for a review see \citealt{Soker2016Rev}).

The cumulative distribution function of the angle $\alpha$ has a very interesting pattern. Below about 40 degrees it follows a perpendicular distribution. This is mainly because objects with $\alpha \la 15^\circ$ are missing. From about 40 degrees to 90 degrees it follows the random distribution.
In any case, the possibility that the NS kick velocity is parallel to the axis of the jets direction is ruled out.

We can think of two basic types of relations between the kick and the jets directions that can explain the missing objects with low values of $\alpha \la 15^{\circ}$. In the first possibility the jets determine the allowed kick direction, while in the second possibility the mechanism that leads to a NS natal kick forces jets in specific directions.

To demonstrate these, we assume that the kick is formed by dense clumps that are formed by instabilities in the ejecta near the NS (e.g., \citealt{Schecketal2006, Wongwathanaratetal2010}). We note that four of the SNRs in our sample (Cassiopeia A, Puppis A, RCW 103, G292.0+1.8) were studied by \cite{Katsudaetal2018} who find that the kick is due to asymmetrical explosion.
The instabilities are likely to result from the standing accretion-shock instability (SASI; see, e.g., \citealt{Abdikamalovetal2015, Fernandez2015, MorenoMendezCantiello2016, Blondinetal2017, Kazeronietal2017}),  or convective overturn that is formed by neutrino heating \citep{Wongwathanaratetal2013}.
One or more dense clumps that are expelled by the explosion, gravitationally attract the NS and accelerate it, in what is termed the gravitational tug-boat mechanism \citep{Janka2017}.
The gravitational tug-boat mechanism is a relatively long-duration process lasting several seconds after accretion has ended, and when the dense regions are accelerated from about 100 km to several thousands km from the origin \citep{Wongwathanaratetal2013, Janka2017}.

\cite{Wongwathanaratetal2010} find (their fig. 2) for their 4 models that the angles between the NS spin and the NS kick are in the range of $\approx 50^\circ -150 ^\circ$. Namely, they are more likely to be perpendicular than aligned.
\cite{Wongwathanaratetal2013} find in their simulations that according to the gravitational tug-boat mechanism in the frame of the delayed neutrino explosion mechanism, there is no correlation between the spin and kick directions.
\cite{Mullleretal2017} obtain similar results. In their simulation the NS spin and NS kick start out as almost perpendicular. After further mass accretion on to the newly born NS the angular momentum axis changes, and the relative angle decreases to $42^\circ$. What they find as the spin of the NS is analog to the general direction of the jets' axis in the jet feedback explosion mechanism.
It is not necessarily the exact jets' axis because the jets might jitter (see section \ref{sec:intro}).

\cite{Wongwathanaratetal2010} and \cite{Wongwathanaratetal2013} also find that in the gravitational tug-boat mechanism in the frame of the delayed neutrino mechanism the NS final velocity is opposite to the direction of the maximum explosion strength. \cite{Janka2017} discusses how the ejection of mass along the polar directions (spin-axis) is delayed, and more mass resides there. As a consequence the kick direction tends to align with the angular momentum axis, but only when a strong spiral SASI mode are present.
In the jet feedback explosion mechanism more mass is concentrated at late times in the equatorial regions, and there is no spin-kick alignment.

Let us then return to the two possibilities within the frame of the jet feedback explosion mechanism, where the angular momentum axis of the accreted gas tends to avoid small angles with respect to the direction of concentration of mass in the instabilities.
In the first possibility the pre-collapse core has a non negligible angular momentum. When it collapses not much material is accreted on to the neutron star from the polar directions \citep{Papishetal2015}. Jets are launched in the general direction of the angular momentum axis. Instabilities can lead to stochastic component of the accreted angular momentum, and the jets might jitter in the vicinity of the angular momentum direction. In any case, the jets further prevent accretion in the vicinity of the polar directions.
Dense clumps will not form close to the polar directions, but rather will tend to form closer to the equatorial plane. Hence, the NS kick will not occur close to the polar directions. The direction of the jets and the direction of the NS natal kick  will avoid each other.

In the second possibility the initial angular momentum does not play a significant role. We start with dense clumps and follow the numerical results of \cite{PapishSoker2014}. When dense clumps are accreted to form an accretion disk, the jets tend to be perpendicular to the accretion direction of dense clumps, and the jets in turn further force accretion perpendicular to their direction of propagation. This behavior leads to a planar jittering-jets pattern \citep{PapishSoker2014}, where the jets' symmetry axes of different jet-launching episodes tend to share the same plane.  Dense clumps tend to form along directions perpendicular to this plane. If the natal kick is caused by dense clumps, this again causes the NS natal kick direction and the direction of jets' axis to avoid each other.

 The real situation might be even more complicated. The `jump' from the perpendicular distribution to the random one comes with concentration of objects, basically two extra objects, around $\alpha=45 ^\circ$. Due to the small number statistics we cannot tell whether this effect is real. It might be, however, a real effect if the missing objects at low values of $\alpha$ are not distributed equally at higher values of $\alpha$, but rather are concentrated on the boundary between the `forbidden' and `allowed' regions of $\alpha$.  

Over all, the jet feedback explosion mechanism might account for the tentative cumulative distribution function for the angle $\alpha$ that we find in the present study.

\section{SUMMARY}
\label{sec:summary}

 We searched the literature for SNRs of CCSNe where we could both identify morphological features, such as ears, that we can attribute to jets and for which the direction of the NS natal kick was determined. We found 12 such SNRs, as we present in Figs. \ref{fig:CassA}-\ref{fig:W44}, and measured the projected (on the plane of the sky) angle between the line connecting the two assumed opposite jets, i.e., the jets' axis, and the NS kick. We summarized the results in Table \ref{Table1}, and plotted the cumulative distribution function (black line) of the angles in Fig. \ref{fig:distribution}. We also plotted there the cumulative distribution functions that are expected from a random distribution (straight orange line) and the distribution expected for a case where the NS kick is always perpendicular to the jets' axis (convex blue line).

 In section \ref{sec:analysis} we compared the cumulative distribution function to the distribution expected from a random distribution and to the distribution expected for a case where the NS kick is always perpendicular to the jets' axis. The cumulative distribution function we find for the 12 SNRs has a $67 \%$ chance to be compatible with the random distribution (straight orange line on Fig. \ref{fig:distribution}), and $12 \%$ to be compatible with the perpendicular distribution (lower convex blue line).
  The basic feature of the cumulative distribution function is that it fits the random distribution at large angles but is missing systems with small angles relative to the random distribution.

 We discussed two possibilities to explain this property, if it is real. Both possibilities assume that dense clumps that are ejected by the explosion accelerate the NS by the gravitational tug-boat mechanism \citep{Wongwathanaratetal2013, Janka2017}, and that jets explode the CCSNe \citep{PapishSoker2011, Soker2016Rev}. Basically, the jets prevent the formation of dense clumps along their propagation direction, or the dense zones supply most of the gas to the accretion disk that launches jets more or less perpendicular to the directions of the dense zones.

 The motivation behind this study is the jet feedback explosion mechanism of massive stars.
 According to the jet feedback explosion mechanism jets that are launched by the newly born NS or black hole drive the explosion of CCSNe. The negative feedback mechanism implies that as long as the jets did not explode the entire core the NS (or black hole if formed) continues to accrete mass from the core. The jets shut themselves off only when they remove the entire core. The last episodes of mass accretion occurs while jets have already expelled the core. Therefore, the last jets that the NS (or black hole) launches expand more freely and can leave an imprint on the ejecta. One of the imprints might be two opposite ears in the SNR \citep{GrichenerSoker2017, Bearetal2017}.

 The main finding of our study is that the jet feedback explosion mechanism, which we consider to be the most promising mechanism to explode all CCSNe, can in principle account for the distribution of angles between the jets' axis and the NS kick velocity.

This research was supported by the Asher Fund for Space Research at the Technion and the Israel Science Foundation.

\end{document}